\documentclass[useAMS,usenatbib]{mn2e}
\usepackage{graphicx}
\usepackage{times}

\begin{document}

\title[Intriguing correlation for the transiting planets]{An intriguing correlation 
between the masses and periods of the transiting planets}

\author[T. Mazeh, S. Zucker \& F. Pont]{Tsevi~Mazeh,$^1$\thanks{E-mail:
mazeh@wise.tau.ac.il} Shay~Zucker$^{2,3}$ and Fr\'ed\'eric~Pont$^2$\\ 
$^1$School of
Physics and Astronomy, Raymond and Beverly Sackler Faculty of Exact
Sciences, Tel Aviv University, Tel Aviv, Israel\\ 
$^2$Observatoire de Gen\`{e}ve, 51 Ch. des Maillettes, Sauverny CH-1290, 
Switzerland\\
$^3$Present Address: Faculty of Physics, Weizmann Institute of Science, 
Rehovot 76100, Israel
}

\maketitle



\begin{abstract}
We point out an intriguing relation between the masses of the
transiting planets and their orbital periods. For the six currently
known transiting planets, the data are consistent with a decreasing
linear relation. The other known short-period planets, discovered
through radial-velocity techniques, seem to agree with this
relation. We briefly speculate about a tentative physical model to
explain such a dependence.
\end{abstract}


\section{Introduction}

The actual masses of most extrasolar planets are not known.  Since
radial-velocity data do not yield orbital inclinations, only
minimum masses can be derived for those planets. Some information
about the masses could be gained through {\it Hipparcos} astrometry
\citep{PouAre2001,ZucMaz2001}, but no definite masses.
So far, the actual masses can be derived only for planets that exhibit
transits, indicating orbital inclinations close to $90\degr$. The
first transiting planet to be discovered was HD\,209458
\citep{Chaetal2000,Henetal2000}, after the radial-velocity modulation
\citep{Mazetal2000} indicated a minimum mass of about $0.7$ Jupiter
Mass (M$_\mathrm{J}$).

The next stage in the pursuit of knowledge of the actual planetary
masses was the publication of the high-quality photometric data of
OGLE \citetext{Udalski et al.\ 2002a,b, 2003},
\nocite{Udaetal2002a,Udaetal2002b,Udaetal2003} which yielded,
especially after applying the BLS transit search algorithm
\citep{Kovetal2002}, more than one hundred transit
candidates. Follow-up radial-velocity observations confirmed that
OGLE-TR-56 \citep{Konetal2003,Toretal2004}, OGLE-TR-113 and
OGLE-TR-132 \citep{Bouetal2004} all have planetary companions, with
masses of $1.45$, $1.35$ and $1.01\,M_\mathrm{J}$, respectively.

In August 2004 three additional steps were taken in the saga of
deriving the planetary masses:
\begin{itemize}
\item Superb photometry of OGLE-TR-132 improved its mass estimate to 
 $1.19\,M_\mathrm{J}$ \citep{Mouetal2004}.
\item One more OGLE candidate, OGLE-TR-111, was proven to harbour a 
planet with a mass of $0.5\,M_\mathrm{J}$ \citep{Ponetal2004}.
\item The first radial-velocity confirmation of planetary transit 
detection by a wide-field small telescope, TrES-1, was announced, with a
mass of $0.75\,M_\mathrm{J}$ \citep{Aloetal2004}.
\end{itemize}
 
\begin{table*}
 \begin{minipage}{240mm}
 \caption{Periods and Masses of the Transiting Planets}
  \begin{tabular}{@{}llll@{}}
\hline
Name         &Period[d] &Mass[$M_J$]        &Reference\\
  \hline
OGLE-TR-56   &   $1.2 $ &  $1.45 \pm 0.23$  &  \citealt{Toretal2004}\\ 
OGLE-TR-113 &    $1.43$ & $1.35 \pm  0.22$  &  \citealt{Bouetal2004}\\ 
OGLE-TR-132  &   $1.69$ & $ 1.19 \pm0.13 $  &  \citealt{Mouetal2004}\\  
TRes-1        &  $3.03$ &  $0.75  \pm 0.06$ &  \citealt{Aloetal2004}\\  
HD 209458     &  $3.52$  & $0.69\pm0.05$    &  \citealt{Mazetal2000}\\  
OGLE-TR-111   &  $4.02$ & $0.53 \pm 0.11$  &   \citealt{Ponetal2004}\\     
\hline
\end{tabular}
\end{minipage}
\label{mass_period_table}
\end{table*}

Table \ref{mass_period_table} summarizes our present knowledge of the
planetary masses known as of 2004 August 31.  In this short
communication we present an intriguing correlation between the masses
of the transiting planets and their periods, and discuss very briefly
its possible implications.

\section{The mass-period correlation}
\begin{figure*}
\includegraphics{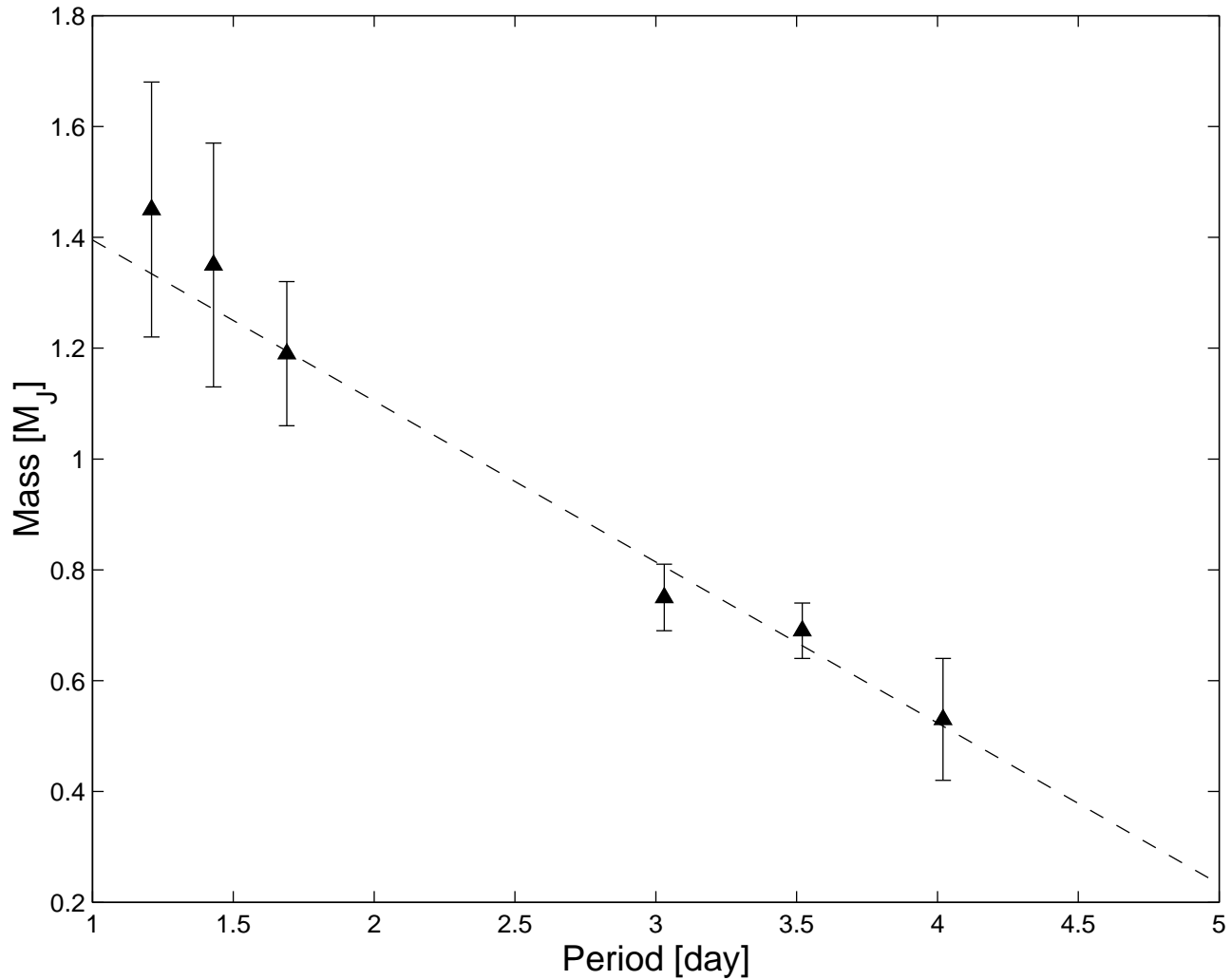} 
\caption{ The masses of the transiting
 planets as a function of their periods .The line is the best linear
 fit.  }
\label{fig:fig1}
\end{figure*}

Fig.\ \ref{fig:fig1} presents the six known planetary masses as a
function of their periods, and their best linear fit. Despite the fact
that we have only six points, the linearity of their positions in the
diagram is intriguing.  The probability to have their masses randomly
arranged as a monotonic (increasing or decreasing) function of their
periods is $2/6!=0.0028$, let alone arrange them on a straight line,
within one sigma.

\begin{figure*}
 \includegraphics{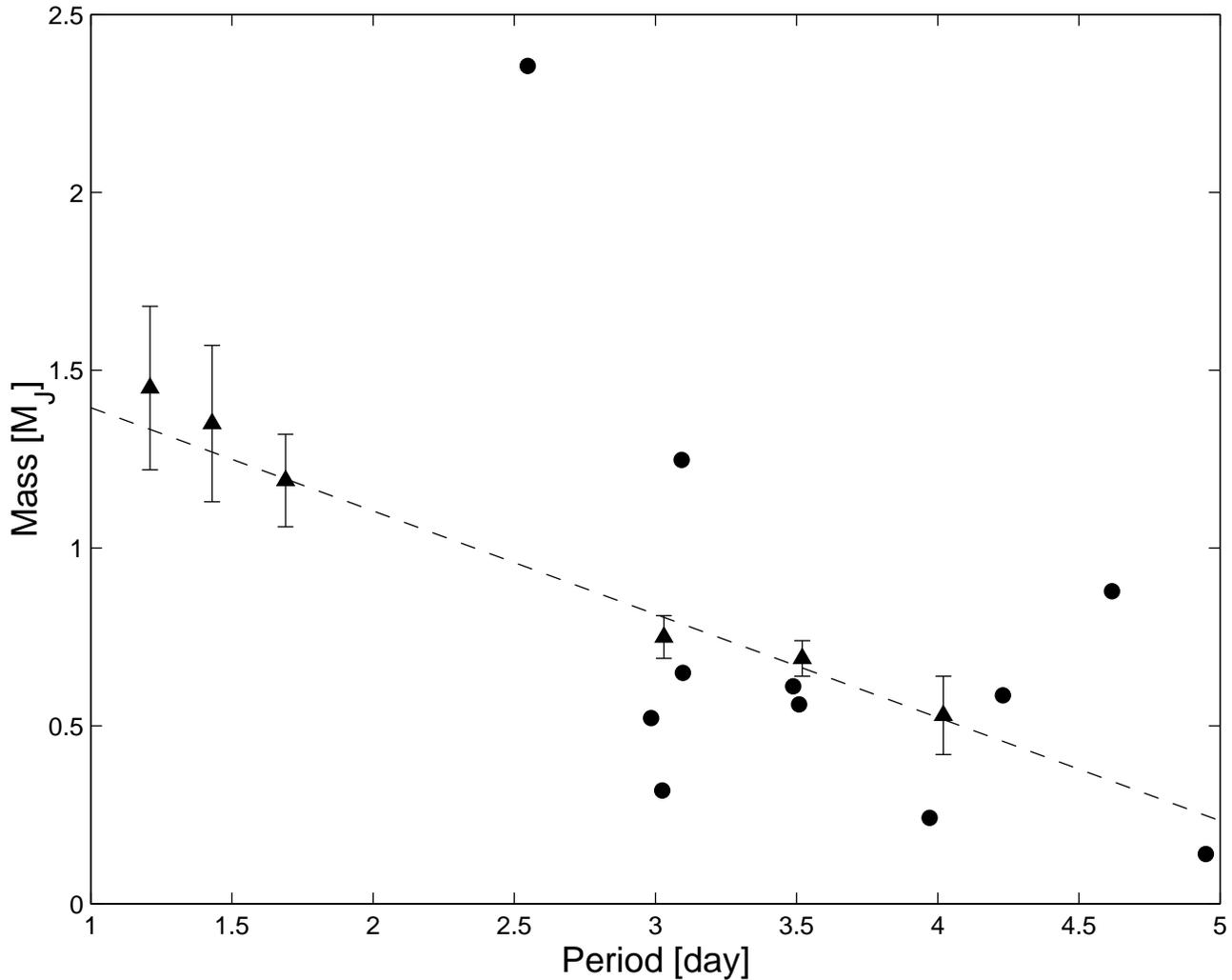} 
 \caption{ The mass-period diagram of the
 short-period planets. The filled circles are the minimum masses
 divided by $\pi /4$. }
\label{fig:fig2}
\end{figure*}

Obviously, any correlation between the masses of the short-period
planets and their periods should also apply to the minimum masses of
the other known short-period planets. We therefore plot in Fig.\
\ref{fig:fig2} the masses of all other planets with periods shorter
than $5$ days (from the Exoplanet Encyclopaedia,
www.obspm.fr/encycl/catalog.html). We exclude only the planet around
$\tau$\,Boo \citep{Butetal1997}, which is known to be in a binary
system \citep{Eggetal2003}, where the distant stellar M2 companion
could have modified the planetary formation and evolution
\citep{ZucMaz2002}. For all these planets only the minimum mass is
known. We therefore divided each planetary minimum mass by $\pi /4$,
the expected value of $\sin\,i$. We ignored the very recently
discovered Neptune-size planets, which probably are of a different
nature and have a different formation and evolutionary history
\citep{Sanetal2004, McAetal2004,Butetal2004}.

We find that the population of all the planets with periods shorter
than $5$ days is still consistent with the intriguing linear
dependence found for the masses of the transiting planets, with
admittedly some scatter.

\citet{ZucMaz2002} have already found a correlation between planet
masses and their orbital periods, manifested as a dearth of massive
short-period planets. The relation suggested here applies to a small
part of the mass-period diagram analyzed by
\citeauthor{ZucMaz2002}. Therefore, it is probably related to other
physical mechanisms.

\section{Discussion}

The suggested mass-period relation depicted in Fig.\ \ref{fig:fig1}
and Fig.\ \ref{fig:fig2} is based on small number statistics, and more
points are certainly needed to establish its existence.  The aim of
the this short note is to attract the attention of the community to
this intriguing relation and to initiate a fruitful discussion. Along
this line, in what follows we speculate very briefly on the possible
mechanism behind such an intriguing feature.

Very recently \citet{Baretal2004} recalculated the evolutionary tracks
of close-in giant planets, taking into account thermal evaporation
caused by the XUV flux of the parent star \citep{Lametal2003}.  They
suggested that the orbital distance determines a critical mass, below
which the evaporation time-scale becomes shorter than the thermal
time-scale of the planet. For planets with initial masses below the
critical mass, evaporation leads to a rapid expansion of the outer
planetary layers, speeding up the evaporation process. Consequently,
planets with masses below the critical mass do not survive. Such a
process can cleanse the area in the mass-period diagram below the line
that represents the critical mass.

Clearly, the critical mass gets smaller as the orbital distance gets
larger. Therefore, such a mechanism can explain the feature seen in
our diagram.  Moreover, planets positioned on the suggested
mass-period line might have suffered an increase of their radius
because of the stellar heating, and would therefore be the easiest to
detect by transit search. This might account for the fact that all the
transiting planets are so close to the line in our diagram.

Obviously this speculation still has to be worked out. For example, it
seems that planets with periods shorter than $5$ days discovered by
radial-velocity measurements also concentrate around our line, and
their density above the line is low. To account for this we need to
invoke another mechanism, as planets above the line apparently are not
strongly affected by stellar heating. Furthermore, one still has to
show why the critical mass would depend linearly on the orbital
period. Moreover, one would think that the evaporation process should
depend strongly on the stellar brightness, a feature which is not seen
in the actual data of the short-period planets. It is also not clear
why this effect is limited to planets with periods shorter than $5$
days. In short, we do not have any detailed model. We suggest that
such a model should be worked out only when more planets in this range
of periods are found, and the mass-period correlation better
established.
  
\section*{Acknowledgments}

This work was supported by the Israeli Science Foundation through
grant no. 03/233. S.Z.\ wishes to acknowledge support by the European
RTN ``The Origin of Planetary Systems'' (PLANETS, contract number
HPRN-CT-2002-00308) in the form of a fellowship. 
\bibliographystyle{mn2e}
\bibliography{ref}


\end{document}